\documentclass[aps,prl,twocolumn,superscriptaddress]{revtex4-2}
\usepackage[utf8]{inputenc}
\usepackage{graphicx}% Include figure files
\usepackage{epstopdf}
\usepackage{bm}% bold math
\usepackage{amsmath}
\usepackage{verbatim}
\usepackage{amssymb}
\usepackage{listings}
\usepackage{color}
\definecolor{darkblue}{rgb}{0,0,0.6}
\usepackage{here}
\usepackage[colorlinks,linkcolor=darkblue,citecolor=darkblue,urlcolor=darkblue]{hyperref}
\DeclareRobustCommand{\average}[1]{\left\langle #1 \right\rangle}
\renewcommand\vec[1]{\boldsymbol{#1}}

\newcommand\fig[1]{Fig.~\ref{#1}}

\renewcommand\vec[1]{\boldsymbol{#1}}
\newcommand\eq[1]{Eq.~(\ref{#1})}

\newcommand\bigpar[1]{\left( #1 \right)}

\begin{document}

\title{Collective relaxation dynamics in a three-dimensional lattice glass model}

\author{Yoshihiko Nishikawa}

\affiliation{Graduate School of Information Sciences, Tohoku University, Sendai 980-8579, Japan}

\author{Ludovic Berthier}

\affiliation{Laboratoire Charles Coulomb (L2C), Universit\'e de Montpellier, CNRS, 34095 Montpellier, France}

\affiliation{Yusuf Hamied Department of Chemistry, University of Cambridge, Lensfield Road, Cambridge CB2 1EW, UK}

\date{\today}

\begin{abstract}
We numerically elucidate the microscopic mechanisms controlling the relaxation dynamics of a three-dimensional lattice glass model that has static properties compatible with the approach to a random first-order transition. At low temperatures, the relaxation is triggered by a small population of particles with low-energy barriers forming mobile clusters. These emerging quasiparticles act as facilitating defects responsible for the spatially heterogeneous dynamics of the system, whose characteristic lengthscales remain strongly coupled to thermodynamic fluctuations. We compare our findings both with existing theoretical models and atomistic simulations of glass-formers.
\end{abstract}

\maketitle

The glass transition of supercooled liquids is characterized by a drastic dynamical slowing down, with timescales that increase much faster than the Arrhenius law in fragile glass-formers~\cite{Debenedetti2001,Berthier2011}. These dynamics become increasingly spatially heterogeneous upon approaching the glass transition~\cite{Ediger2000,DynamicalHeterogeneities2011}, in a manner revealed by four-point dynamic susceptibilities and correlation functions~\cite{lacevic2003spatially,DynamicalHeterogeneities2011,berthier2005dynamical,karmakar2014growing}. Over the last decade, massive efforts were also deployed to establish and characterize the growth of relevant thermodynamic fluctuations and static correlation lengthscales accompanying this slowing down~\cite{biroli2008thermodynamic,karmakar2009growing,Kob2012,berthier2013overlap}, with results compatible with an approach to a random-first order transition (RFOT)~\cite{Berthier2017,berthier2019zero,guiselin2020random,guiselin2022statistical,guiselin2022glass}.

The central open question that remains, then, concerns the existence of a causal link between static and dynamic fluctuations, to go beyond known correlations~\cite{richert1998dynamics}. While RFOT theory suggests that static fluctuations directly control the evolution of dynamic heterogeneity and explain the slowing down of the dynamics~\cite{Kirkpatrick1989,Bouchaud2004,biroli2022rfot}, alternative approaches use local energy barriers and dynamic facilitation~\cite{Garrahan2002,Garrahan2003,Chandler2010,Keys2011,dyre2006colloquium}, making no reference to the underlying complex free energy landscape.

Experimental measurements do not easily discriminate these views~\cite{albert2016fifth}, and original strategies are needed~\cite{berthier2021self}. Recent developments in computational studies of atomistic glass-formers~\cite{berthier2023modern} helped reveal the importance of dynamic facilitation at very low temperatures~\cite{Chacko2021,Guiselin2022,Nishikawa2022,Scalliet2022}, strengthening earlier results~\cite{vogel2004spatially,Keys2011}. The reported decoupling between static and dynamic lengthscales~\cite{charbonneau2013decorrelation,Scalliet2022} weakens the idea that statics fully controls dynamics, as does the observation that local Monte Carlo algorithms strongly impact the equilibrium dynamics~\cite{Wyart2017}. While RFOT theory accounts for all these observations~\cite{xia2001microscopic,biroli2022rfot,Berthier2019cantheglass}, the current situation is confused.

Simplified lattice models with minimal (but not too minimal) ingredients are ideally suited to clarify these questions~\cite{Berthier2011}. Kinetically constrained lattice glass models have no complex thermodynamics but display dynamic heterogeneity stemming instead from dynamic facilitation and kinetic constraints~\cite{Fredrickson1984,jackle1991hierarchically,Kob1993,Toninelli2006,ritort2003glassy}. Interacting plaquette spin models form another class, in which similar kinetic constraints emerge from interacting degrees of freedom~\cite{garrahan2002glassiness,Jack2005,Jack2016}. Finally, lattice glass models with frustrated interactions also exist~\cite{Biroli2001,Ciamarra2003,rivoire2004glass,Darst2010,Foini2011,Ciamarra2003,Darst2010}. Differently from the other two families, they display, just like atomistic models, a random first-order transition in the mean-field limit. They are therefore ideal candidates to study the interplay between static and dynamic fluctuations in finite dimensions. This is the central motivation for our work.

Recently~\cite{Nishikawa2020}, a lattice glass model with good glass-forming ability in $d=3$ was numerically shown to display thermodynamic fluctuations consistent with both RFOT theory and finite-$d$ glass-formers. Here, we study its dynamics to shed light on how structural relaxation occurs in a many-body particle system approaching a RFOT. We find that the dynamics becomes glassy and spatially heterogeneous, with a strong coupling between static and dynamic fluctuations. However, structural relaxation is triggered by emerging quasiparticles composed of localized clusters of particles with low-energy barriers, which gradually relax the correlated slow regions in a manner reminiscent of dynamic facilitation. While statics plays a role in the dynamics, this differs strongly from the predictions of RFOT theory. 

We study a binary mixture of particles~\cite{Nishikawa2020} on a periodic cubic lattice of linear dimension $L$ and Hamiltonian
\begin{equation}
H = \sum_{i} \bigg( \sum_{j} 
\delta(|\vec r_i - \vec r_j|, 1) - \ell_{\sigma_i} \bigg)^2,
\label{eq:hamiltonian}
\end{equation}
with $\delta(\cdot, \cdot)$ the Kronecker delta, $\vec r_i$ the position of particle $i$ taking values $r_{i, \alpha} = 1, 2, \cdots L$ ($\alpha = x, y, z$), and $\sigma_i \in \{1, 2\}$ specifying the particle type. Each particle type has a preferred number of neighbors, $\ell_\sigma$, and quadratic deviations from this number provide the local energy cost, see SM~\cite{Supplement}. We set $\ell_1 = 3$ and $\ell_2 = 5$, fix the density to $\phi = N / L^3 = 0.75$, where $N$ is the number of particles, and the concentrations $\rho_1 = N_1 / N = 0.4$ and $\rho_2 = N_2/N=0.6$~\cite{Nishikawa2020}. We use $L = 20$, apart from the temperature $T = 0.26$, where we set $L = 10$ as the simulations take much longer to converge.

To obtain equilibrium configuration of the system defined by \eq{eq:hamiltonian} we use a non-local swap dynamics, where two lattice sites with different occupation numbers or particle types are randomly chosen and swapped with Metropolis probability. To study the physical dynamics, we restrict the above rule to pairs of nearest neighbors~\cite{Nishikawa2020}. This local dynamics allows us to explore structural relaxation down to $T \approx 0.3$, while Ref.~\cite{Nishikawa2020} locates the putative Kauzmann transition near $T \approx 0.24$. At $T=0.3$, the non-local swap provides a comfortable speedup of about $10^3$, allowing us to explore the equilibrium dynamics of the model at temperature where the physical dynamics is extremely slow. The unit time represents $L^3$ attempted moves. 

\begin{figure}
\includegraphics[width=\linewidth]{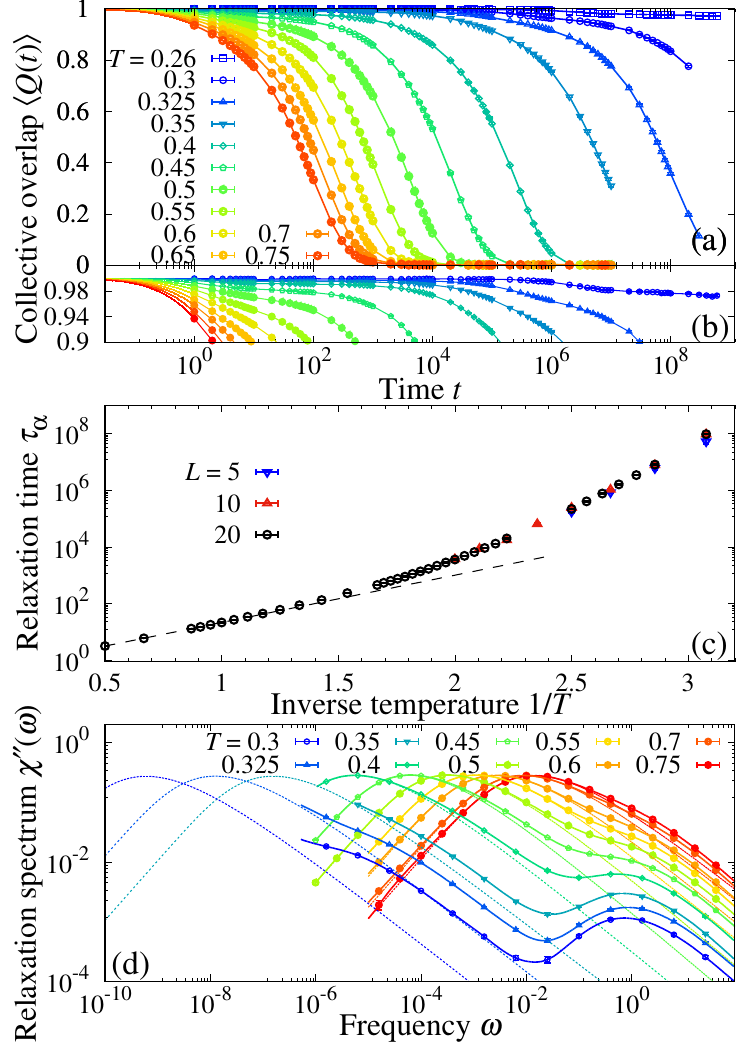}
\caption{(a, b) Temperature evolution of the collective overlap $\average{Q(t)}$.
(c) Arrhenius plot of the relaxation time $\tau_\alpha$; the broken line is an Arrhenius fit valid at high temperatures.
(d) Relaxation spectrum from \eq{eq:spectrum}; dashed curves represent the spectrum corresponding to the final stretched exponential decay of the overlap.}
\label{fig:overlap_tau_alpha}
\end{figure}

The structural relaxation of the system can be followed using the collective overlap
\begin{equation}
\average{Q(t)} = \frac{1}{1-Q_0} \average{\frac1N \sum_i q_i(t) - Q_0}.
\label{eq:collectiveQ}
\end{equation}
In this expression the local overlap $q_i(t)$ remains $1$ only when the site occupied by particle $i$ at time $0$ is again occupied by any particle of the same type at time $t$ later, where $q_i(t) = \sum_j \delta(\vec r_i(0), \vec r_j(t))\delta(\sigma_i(0), \sigma_j(t))$ and $Q_0 = \phi(\rho_1^2 + \rho_2^2)$. The brackets $\average{\cdot}$ stand for an average over initial equilibrium configurations. 
The collective overlap in Eq.~(\ref{eq:collectiveQ}) is the lattice analog of the overlap defined for particle systems~\cite{Kob2012}. It is normalized to decay from 1 to 0 when the structure at $t=0$ has fully relaxed. Numerical results are shown in Fig.~\ref{fig:overlap_tau_alpha}(a). At high temperatures $T \gg 1$, the system rapidly decorrelates. When $T \lesssim 0.55$, $\average{Q(t)}$ starts to have a plateau [see zoom in \fig{fig:overlap_tau_alpha}(b)] and the corresponding relaxation time $\tau_\alpha$ departs from an Arrhenius behavior, see \fig{fig:overlap_tau_alpha}(c).

With decreasing $T$ further, the dynamics keeps slowing down, and at very low temperatures $T \lesssim 0.325$, a more complex short time dynamics emerges. This is better appreciated in the Fourier relaxation spectrum~\cite{berthier2005numerical}
\begin{equation}
    \chi^{\prime\prime}(\omega) = 
    \int d\log\tau \frac{d\average{Q(\tau)}}{d\log(\tau)}\frac{\omega \tau}{1 + (\omega\tau)^2},
\label{eq:spectrum}
\end{equation}
as shown in \fig{fig:overlap_tau_alpha}(d) (see details in SM~\cite{Supplement}). This representation clearly reveals additional relaxation processes taking place at intermediate frequencies between the microscopic peak and the main peak corresponding to the structural relaxation, in qualitative agreement 
with recent observations in particle models~\cite{Guiselin2022,Nishikawa2022}.

\begin{figure}
\includegraphics[width=\linewidth]{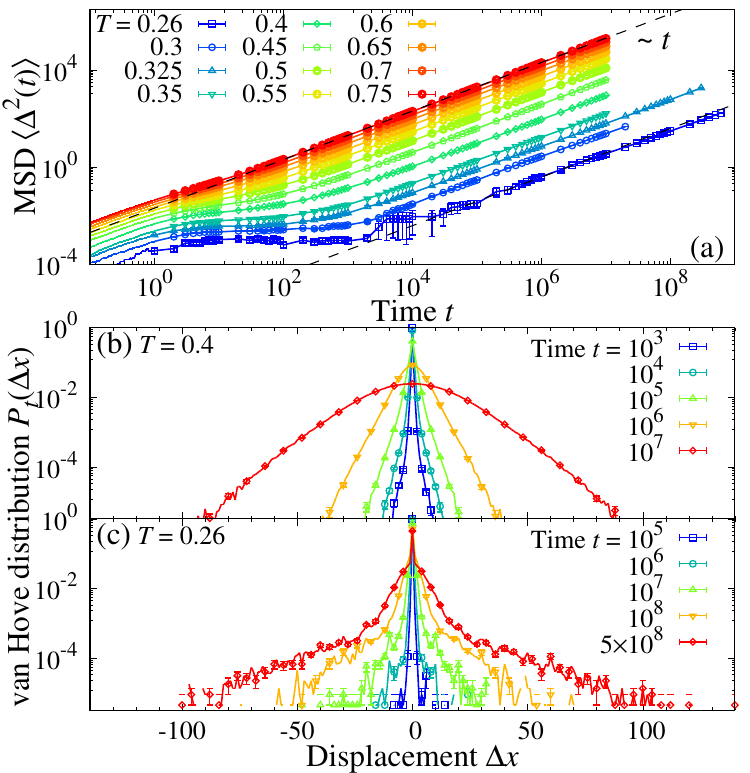}
\caption{(a) Time dependence of the MSD; diffusive behavior is indicated with dashed lines. Van Hove distribution functions at (b) $T=0.4$ where 
$\tau_\alpha \simeq 2\times 10^5$ and (c) $T = 0.26$ where 
$\tau_\alpha \gg 10^{10}$.}
\label{fig:MSD}
\end{figure}

We turn to the mean-squared displacement (MSD)
\begin{equation}
    \average{\Delta^2(t)} = \average{\frac1N\sum_i \bigl|\vec r_i(t) - \vec r_i(0)\bigr|^2}, 
\end{equation}
to understand how particle motion leads to structural relaxation, see \fig{fig:MSD}(a). We find, as usual, that the MSD only becomes diffusive at long enough times, after a transient plateau that becomes longer at lower temperatures and a decreasing diffusion constant. However, the MSD slows down much less than the overlap. For instance, at the lowest temperature $T=0.26$ and at $t=3\times 10^8$, $\average{Q(t)} \simeq 0.98$ while $\average{\Delta^2(t)} \simeq 10^2$. It is a priori surprising that such large particle displacements lead to so little relaxation in the structure. 

To better understand this finding, we determine the corresponding van Hove distribution
\begin{equation}
P_t(\Delta x) = \average{\frac 1{3N}\sum_{i,\alpha} \delta\bigpar{\Delta x - (r_{i,\alpha}(t) - r_{i,\alpha}(0))}},
\label{eq:vanHove}
\end{equation}
see \fig{fig:MSD}(b,c). At high temperatures, the particle dynamics is diffusive and $P_t(\Delta x)$ very close to a Gaussian at any time. In contrast, $P_t(\Delta x)$ is non-Gaussian at short times with nearly-exponential tails~\cite{Chaudhuri2007} when decreasing temperature, and only becomes Gaussian when $t \gg \tau_\alpha$, see $T=0.4$ in \fig{fig:MSD}(b). At very low temperature, e.g., $T=0.26$ in \fig{fig:MSD}(c), $P_t(\Delta x)$ develops extremely extended tails coexisting with a large peak at $\Delta x = 0$. This peak corresponds to a large fraction of particles which have not moved at all since $t=0$, coexisting with a population of particles which have covered distances up to tens of lattice sites. Clearly, dynamics is highly heterogeneous, and different particles can exhibit different behavior. Also, the large decoupling between the overlap and the MSD arises because a small fraction of particles travels large distances while many others have not yet moved. 

\begin{figure}
\includegraphics[width=\linewidth]{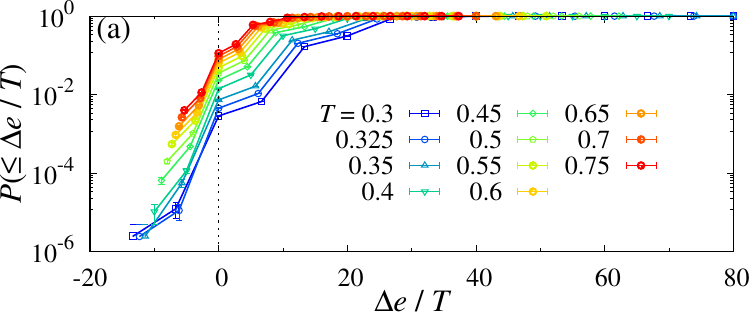}
\includegraphics[width=\linewidth]{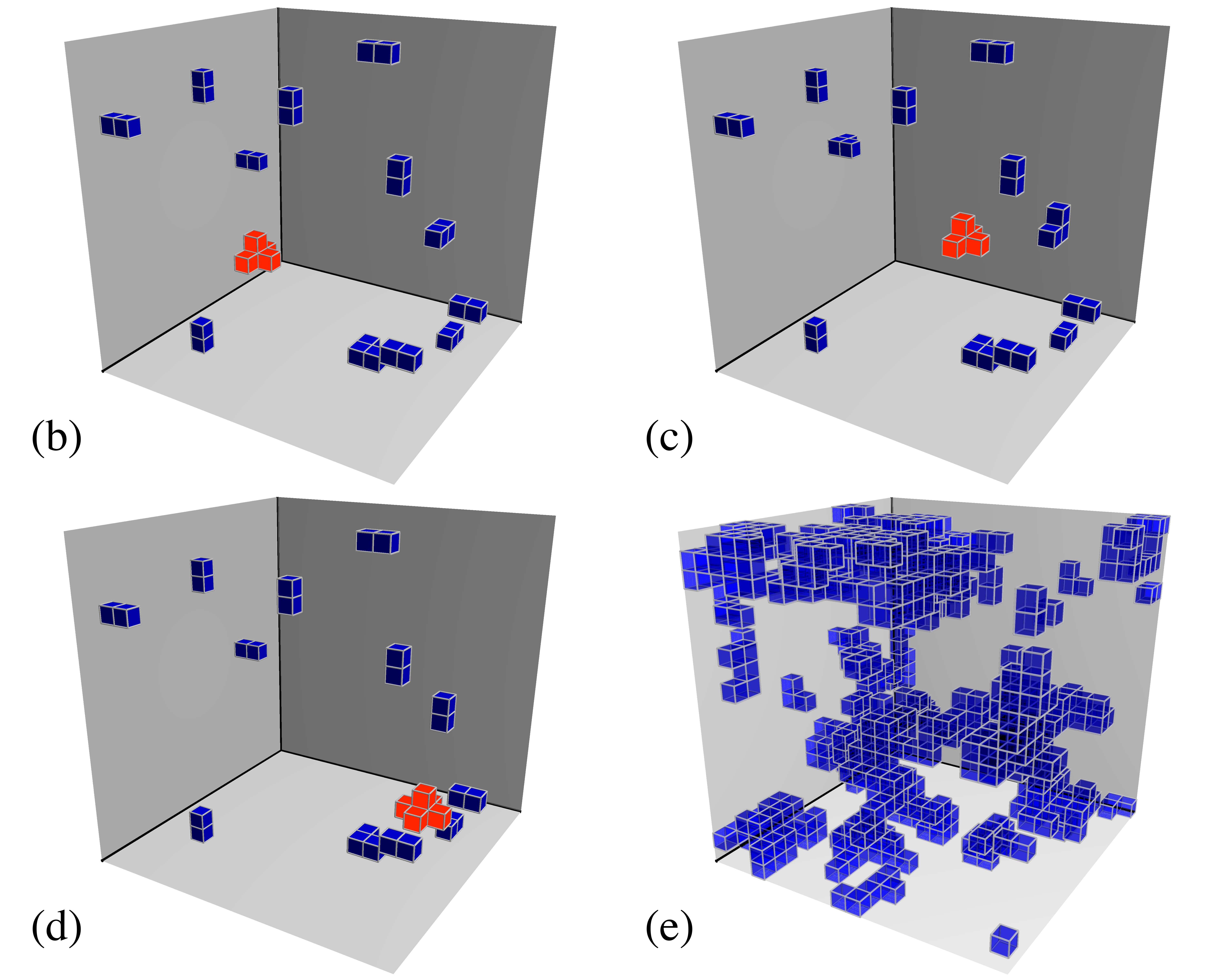}
\caption{(a) Cumulative distribution of local energy barriers. The vertical line indicates $\Delta e=0$. 
(b-d) Snapshots showing lattice sites with $\Delta e \leq 0$ at $T = 0.3$ at three different times. The time intervals between (b) and (c) and between (c) and (d) are $50$ and $350$, respectively. Sites highlighted in red belong to a quasiparticle moving rapidly. (e) Lattice sites with states that differ between times $0$ and $7\times 10^4$, showing the superposition of multiple quasiparticle paths.}
\label{fig:energy_barrier}
\end{figure}

The emergence of a small population of fast moving particles in an otherwise nearly frozen backbone is unexpected as it is not directly included in the Hamiltonian in \eq{eq:hamiltonian}, contrary to kinetically constrained models. To understand this feature, we measure for lattice site $i$ in an equilibrium configuration a local energy barrier, $\Delta e_i$, defined as the minimum energy cost to swap with one of its nearest neighbors. Such analysis is not possible in off-lattice models, but is very easy here. We show in \fig{fig:energy_barrier}(a) the cumulative probability distribution of $\Delta e / T$. Lattice sites with $\Delta e / T \geq 10$, for instance, are only swapped with probability $\leq 10^{-4}$. At high temperature, a large fraction of lattice sites have $\Delta e / T \leq 0$ and can move with no rejection. With decreasing temperature, local energy barriers become very large. At $T = 0.3$, a majority of sites have $\Delta e / T \geq 26$, yielding a local timescale $\geq 10^{11}$. Strikingly, however, a fraction of about $0.1\%$ of the lattice sites remains totally free to swap. 

In \fig{fig:energy_barrier}(b-d), we show the lattice sites with $\Delta e \leq 0$ at $T=0.3$. A significant fraction of these are dimers and trimers, which result in reversible local particle exchanges. More rarely, we observe a larger cluster, as highlighted in red. Our visualizations indicate that these localized clusters can move very rapidly throughout the system. In the example of \fig{fig:energy_barrier}, the cluster travels more than $L/2$ over about $400$ time steps. The identity of the particles that belong to the cluster also changes rapidly. Therefore, these fast moving localized clusters are {\it emerging quasiparticles} which can move large distances. Particles advected by clusters are responsible for the extended tails in the van Hove distributions. 
After many quasiparticles have been observed, the sites that have relaxed are not homogeneously distributed, as shown in \fig{fig:energy_barrier}(e). This suggests the existence of specific paths along which the motion of quasiparticles occurs preferentially. As a corollary, there exist large compact domains in which fast motion cannot occur and can only relax by the repeated motion of many quasiparticles at their boundaries. This represents, for the lattice glass model under study, the analog of the dynamic facilitation reported for off-lattice simulations~\cite{Keys2011,Scalliet2022}. 
 
\begin{figure}
\includegraphics[width=\linewidth]{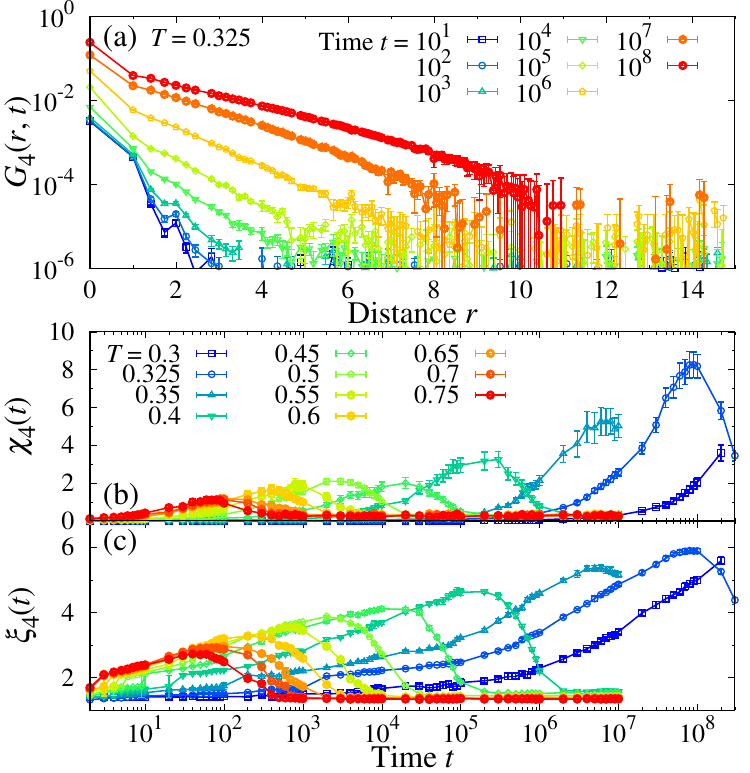}
\caption{(a) Four-point correlation function $G_4(r, t)$ at $T = 0.325$ where $\tau_\alpha \approx 10^8$.  
(b) Dynamical susceptibility $\chi_4(t)$ and (c) dynamical lengthscale $\xi_4(t)$ demonstrating increasing dynamic heterogeneity.}
\label{fig:G4}
\end{figure}

To describe more quantitatively the emerging dynamic heterogeneity, we measure the four-point correlation function $G_4(r, t)$, defined as  
\begin{equation}
    G_4(r,t) = \frac1{4\pi r^2}\average{\frac1N\sum_{i, j} \delta q_i(t) \delta q_j(t) \delta(r - |\vec r_i - \vec r_j|)},
\end{equation}
with $\delta q_i(t) = q_i(t) - \sum_j\average{q_j(t)}/N$ the local fluctuation of the overlap. We extract the dynamic lengthscale $\xi_4(t)$ using the definition $G_4(\xi_4, t) / G_4(0, t) = 10^{-2}$, as well as the dynamical susceptibility $\chi_4(t) = \int dr 4\pi r^2 G_4(r, t)$. At high temperatures, $G_4(r, t)$ decays rapidly over a few lattice sites even near $\tau_\alpha$, showing that relaxation is purely local. Spatial correlations revealed by a much slower decay in $G_4(r, t)$ appear at lower temperatures, see \fig{fig:G4}(a) for $T=0.325$ where $\xi_4$ grows up to $\xi_4 \approx 6$ near $\tau_\alpha \approx 10^8$. This evolution is observed for both $\xi_4$ and $\chi_4$, whose time dependencies are shown in \fig{fig:G4}. These data are familiar~\cite{berthier2005dynamical}, with a slow growth for $t \ll \tau_\alpha$ followed by a maximum near $\tau_\alpha$. Interestingly, at very low temperatures, the slow growth of $\xi_4$ towards its peak is compatible with a power law, $\xi_4 \sim t^{1/z}$, with a small exponent $1/z \approx 0.15$, in line with recent off-lattice findings~\cite{Scalliet2022}. 

\begin{figure}
\includegraphics[width=\linewidth]{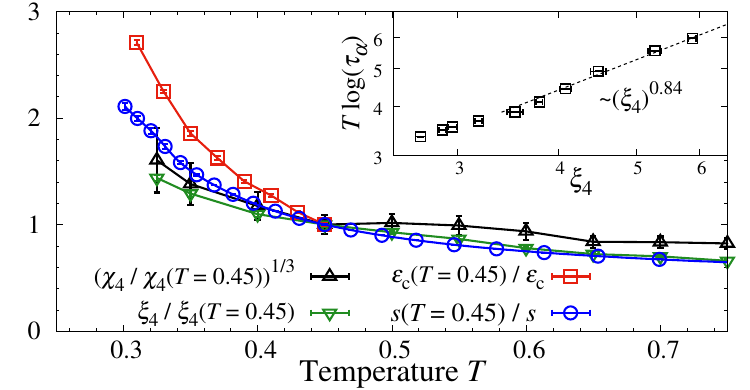}
\caption{Evolution of two estimates of a dynamic lengthscale, $(\chi_4)^{1/3}$ and $\xi_4$, and of two estimates of a static correlation lengthscale $1/s(T)$ and $1/\varepsilon_c(T)$. Each quantity is normalized by its value at $T=0.45$. Inset shows $T\log(\tau_\alpha)$ as a function of $\xi_4$, the dashed line is a fit to $\log(\tau_\alpha) \sim \xi_4^\psi/T$ with $\psi = 0.84$.}
\label{fig:lengths}
\end{figure}

We collect all relevant lengthscales in \fig{fig:lengths} where we show both $\xi_4$ and $\chi_4$ measured at their maximum near $\tau_\alpha$. Our data indicate that the simple relation $\chi_4 \sim \xi_4^3$ holds (not shown), confirming that slow domains have a compact geometry and thus we report the quantity $\chi_4^{1/3}$ in \fig{fig:lengths}. We normalize these quantities by their values at temperature $T=0.45$, below which important thermodynamic quantities have been determined before~\cite{Nishikawa2020}. 

Two estimates of the configurational entropy~\cite{Berthier2019} were measured. First, the total entropy $s(T)$ itself was measured using thermodynamic integration. It should be close to the configurational entropy since vibrational contributions are negligible on the lattice. Second, the inverse of the critical coupling field $\varepsilon_c$~\cite{Nishikawa2020} in the Franz-Parisi scheme~\cite{Franz1997} also represents a solid estimate of the configurational entropy~\cite{Berthier2014}, which is itself inversely proportional to the point-to-set correlation lengthscale quantifying static correlations~\cite{Kirkpatrick1989,Bouchaud2004,Berthier2019} {(see SM~\cite{Supplement} for a review of these connections).} These two indirect estimates of a static lengthscale are included in \fig{fig:lengths}, also rescaled at $T=0.45$. They appear to grow at least as significantly (if not more strongly) than dynamic lengthscales. The coupling between static and dynamic lengthscales is therefore stronger here than in atomistic models~\cite{Kob2012,Berthier2012,charbonneau2013decorrelation,Scalliet2022}. This finding encourages us to directly test the relation between timescales and lengthscales predicted in RFOT theory, see \fig{fig:lengths}, where the relation $\log \tau_\alpha \sim \xi_4^\psi / T$ with $\psi \approx 0.84$ is followed by the data. While a larger exponent $\psi$ (between $d/2$ and $d$) is expected from cooperative relaxation events, it is not surprising that a lower apparent value is found here, as dynamic facilitation via mobile quasiparticles provides more efficient dynamic pathways. {This result indicates that the RFOT theory picture of cooperative activated dynamics is not dominant for the studied model.} Interestingly, values $\psi<1$ were also inferred from analysis of experimental data~\cite{capaccioli2008dynamically,ozawa2019does} and numerically~\cite{karmakar2009growing,cammarota2009numerical}.   

In conclusion, our results establish that, differently from all other classes of lattice models for glassy dynamics, the $3d$ lattice glass model studied here displays thermodynamic and dynamic properties that compare favorably with atomistic glass-formers, with growing static point-to-set and four-point dynamic lengthscales. While static and dynamic fluctuations appear to be strongly coupled, the relaxation dynamics is nevertheless different from the RFOT theory description invoking cooperative relaxation events. Instead, while the system genuinely approaches a random first-order transition, it also leaves behind a small population of weakly constrained particles forming emergent quasiparticles whose fast propagation slowly relaxes the entire system eventually at large times. These quasiparticles thus resemble the dynamic defects {postulated in kinetically constrained models,} or those emerging from the specific interactions of plaquette models. They also lead to signatures at intermediate timescales that are reminiscent of recent findings in atomistic models~\cite{Scalliet2022,Guiselin2022}. 

{While details of the emergent quasiparticles and dynamical defects may be specific to our model, the physical picture emerging from our study appears generic. It is consistent with the recent body of results regarding glassy dynamics at very low temperatures and it supports the generic conclusion that cooperative events from statically correlated domains are preempted by faster relaxation events involving localized regions acting as facilitating defects \cite{Mishra2014,Scalliet2022,Nishikawa2022}. Our results provide a unified picture for glassy dynamics near the glass transition where facilitated dynamics takes place near localized regions, whose temperature evolution is strongly connected to emerging static correlations. We hope that future studies will confirm the validity of these conclusions across a broader range of models (including molecular systems) and spatial dimensions, and study the fate of models even closer to the putative ideal glass transition where cooperative events could finally become dynamically relevant.}  

\acknowledgments

We thank P.~Charbonneau, A.~Ikeda, H.~Ikeda, J.~Takahashi, H.~Yoshino, and F.~Zamponi for useful discussions. Y.N. acknowledges support from JSPS KAKENHI (Grant No.~22K13968), L.B.'s work is supported by a grant from the Simons Foundation (\#454933).  

\bibliography{refs}

\end{document}

\title{Supplemental Material for ``Collective relaxation dynamics in a three-dimensional lattice glass model''}

\author{Yoshihiko Nishikawa}

\affiliation{Graduate School of Information Sciences, Tohoku University, Sendai 980-8579, Japan}

\author{Ludovic Berthier}

\affiliation{Laboratoire Charles Coulomb (L2C), Universit\'e de Montpellier, CNRS, 34095 Montpellier, France}

\affiliation{Yusuf Hamied Department of Chemistry, University of Cambridge, Lensfield Road, Cambridge CB2 1EW, UK}

\maketitle

\subsection{Model}

We describe the lattice glass model in more detail. The Hamiltonian of the model is 
\begin{equation}
H = \sum_{i} \bigg( \sum_{j} \delta(|\vec r_i - \vec r_j|, 1) - \ell_{\sigma_i} \bigg)^2,
\end{equation}
where $\delta(\cdot)$ is the Kronecker delta. Each particle lies on a lattice site, and each site cannot be occupied by more than one particle. 

The energy of particle $i$ depends on the number of neighboring particles, which is counted in the term $\sum_{j} \delta(|\vec r_i - \vec r_j|, 1)$, and on the parameter $\ell_{\sigma_i}$ with $\sigma_i \in \{1, 2\}$ denoting the particle type. The energy for particle $i$ is zero only when $\sum_{j} \delta(|\vec r_i - \vec r_j|, 1) = \ell_i$, and it is positive when the equality is not satisfied. Therefore it is energetically favorable that particles surround themselves by exactly $\ell_i$ neighbors, but simultaneously satisfying this constraint for all particles is not possible. In our study, we use a binary mixture of particles with $\ell_1 = 3$ or $\ell_2 = 5$, and we can think of particles with a smaller $\ell$ as being ``larger'' than particles with a larger $\ell$ which can possess more neighbors.

We mention the relation with another lattice glass model proposed by Biroli and M\'ezard (BM) \cite{Biroli2001} and its generalization \cite{Foini2011}. 
The original BM model has a hard-core interaction, where, for particle $i$ with a parameter $\ell_i$, the number of neighboring particles cannot exceed $\ell_i$: The energy cost becomes infinity when the number of neighbors becomes larger than $\ell_i$, and it is zero otherwise. Ref.~\cite{Foini2011} then introduced a generalized BM model with the Hamiltonian
\begin{equation}
H = \sum_{i} \bigg( \sum_{j} \delta(|\vec r_i - \vec r_j|, 1) - \ell_{\sigma_i} \bigg)\theta\bigg( \sum_{j} \delta(|\vec r_i - \vec r_j|, 1) - \ell_{\sigma_i} \bigg),
\end{equation}
where $\theta(\cdot)$ is the Heaviside step function. In the zero-temperature limit, this generalized model converges to the original, hard-core BM model. 
Comparing the generalized BM model and our lattice glass model, we notice that the number of neighbors achieving the minimum energy differs in the models: In our lattice glass model, only when $\sum_{j} \delta(|\vec r_i - \vec r_j|, 1) = \ell_i$ the interaction energy is zero while, in the generalized BM model, any number of neighbors smaller than or equal to $\ell_i$ yields zero energy. While all three models only differ by the specific choice of the constraints, it results in very different dynamic and static behaviors at low temperature and high density.

\subsection{The relaxation spectrum}

The relaxation spectrum $\chi^{\prime\prime}(\omega)$ is given by
\begin{equation}
    \chi^{\prime\prime}(\omega) = 
    \int d\log\tau \frac{d\average{Q(\tau)}}{d\log(\tau)}\frac{\omega \tau}{1 + (\omega\tau)^2}.
\end{equation}
To compute the spectrum~\cite{Guiselin2022}, we first interpolate $\average{Q(t)}$ on a very fine temporal grid, and we then numerically differentiate the interpolated $\average{Q(t)}$, to perform the numerical integration of $\frac{d\average{Q(\tau)}}{d\log(\tau)}\frac{\omega \tau}{1 + (\omega\tau)^2}$. 

\begin{figure}
    \centering
    \includegraphics[width=.6\linewidth]{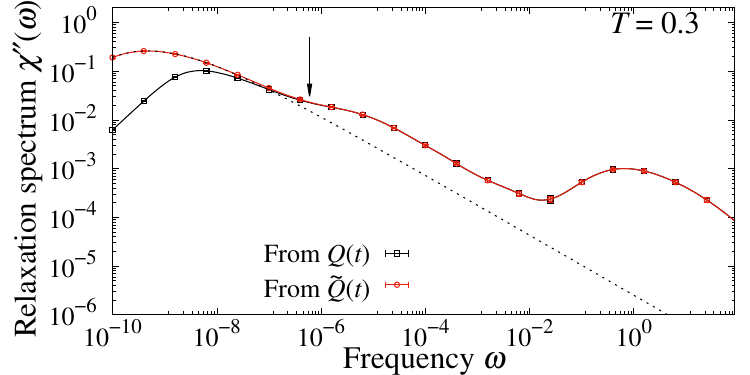}
    \caption{Relaxation spectra of the lattice glass model at $T=0.3$ from the real and synthetic data. At $\omega \gtrsim 10^{-6}$, two spectra are almost identical while at smaller $\omega$, they significantly deviate from each other due to the limited simulation time. In the main figure, we only show the real data when they coincide with the synthetic one. The broken line represents the spectrum of the stretched exponential decay fitted to the observed overlap $\average{Q(t)}$ and the arrow indicates the smallest $\omega$ for $\chi^{\prime\prime}(\omega)$ we show in the main figure.}
    \label{fig:spectrum_T03}
\end{figure}

However, at low temperatures, e.g., $T = 0.3$, $0.325$, and $0.35$, we have not reached a timescale where $\average{Q(t)} \simeq 0$ due to extremely long relaxation times, which could result in systematic errors in $\chi^{\prime\prime}(\omega)$ at small $\omega$. 
To avoid the errors, we compare $\chi^{\prime\prime}(\omega)$ from $\average{Q(t)}$ and the one from an extrapolated $\average{Q(t)}$ as follows.
We first fit a stretched exponential function $\sim \exp(-(t / \tau)^\beta)$ to the observed collective overlap using only data points where $\average{Q(t)} < 0.9$. We then construct an extrapolated overlap $\average{\tilde Q(t)}$ that decays to $0$ at a very long time and compute the relaxation spectrum $\tilde{\chi}^{\prime\prime}(\omega)$ from it. For a given $\omega$, if $\tilde{\chi}^{\prime\prime}(\omega) \simeq \chi^{\prime\prime}(\omega)$, then the extrapolation does not affect the relaxation spectrum, and we can then safely say that the systematic error due to a limited simulation time is negligible there, see \fig{fig:spectrum_T03} for the spectra at $T=0.3$. In Fig.~1(d), we show the relaxation spectrum only for the $\omega$ regime where we have confirmed $\left|\chi^{\prime\prime}(\omega) - \tilde{\chi}^{\prime\prime}(\omega)\right| / \chi^{\prime\prime}(\omega) < 2 \times 10^{-2}$, which implies that truncation of the Fourier integral contributes very little to the results. 

\subsection{The Franz-Parisi potential and the $\varepsilon$-coupling construction}

We consider the system coupled to a fixed configuration $\vec r^\prime$ sampled at inverse temperature $\beta$, as originally proposed in Ref.~\cite{Franz1997}. The total Hamiltonian of this coupled system is
\begin{equation}
    H_\varepsilon(\vec r | \vec r^\prime) = H(\vec r) - \varepsilon Q(\vec r, \vec r^\prime).
\end{equation}
where $Q(\vec r, \vec r^\prime)$ is the instantaneous collective overlap between two configurations $\vec r$ and $\vec r^\prime$, 
\begin{equation}
    Q(\vec r, \vec r^\prime) = \frac1{1 - Q_0} \left( \frac1N\sum_{i, j} \delta(\vec r_i, \vec r^\prime_j)\delta(\sigma_i, \sigma^\prime_j) - Q_0\right).
\end{equation}
Coupling to a fixed configuration introduces quenched disorder into the system, and the system can have a finite overlap with the fixed configuration. We measure the overlap distribution averaged over fixed configurations,
\begin{equation}
    [P_\varepsilon(q)] = [\average{\delta(q - Q(\vec r, \vec r^\prime))}_\varepsilon],
\end{equation}
where the brackets $[\cdot]$ stands for an average over fixed configurations and $\average{\cdot}_\varepsilon$ is the thermal average of the coupled system. We then calculate the overlap susceptibility
\begin{equation}
    \chi_\text{SG} = N \left[ \average{q^2}_\varepsilon - \average{q}_\varepsilon^2 \right]
\end{equation}
as a function of the coupling $\varepsilon$. At low temperature, Ref.~\cite{Nishikawa2020} shows that the coupling induces a first-order phase transition at $\varepsilon_c$, where the susceptibility $\chi_\text{SG}$ diverges in the limit $N\to\infty$. We thus estimate the transition point $\varepsilon_c$ as the peak position in $\chi_\text{SG}$ of the system with $L=12$.

The Franz-Parisi potential $V(Q)$ is related to the probability distribution of the overlap, $ V(Q) \sim  -(\beta N)^{-1} \log  [P(Q)]  ] $ and represents the free energy cost to observe the value $Q$ of the overlap. The free energy difference between large and small $Q$ values is proportional to $\varepsilon_c$. It is also an estimate of the configurational entropy $s_\text{conf}$~\cite{Berthier2014} and the point-to-set static lengthscale $\xi_\text{PTS}$~\cite{Berthier2019} with a nontrivial surface exponent $\theta < d$ as
\begin{equation}
    \varepsilon_c \sim s_\text{conf} \sim \xi_\text{PTS}^{\theta - d}.
\end{equation}

\bibliography{refs}